\documentclass{aastex}

\slugcomment{ Accepted 9/9/99 for publication in { \it The Astrophysical
Journal.} }

\shortauthors{Rubenstein \& Schaefer}
\shorttitle{Superflares caused by extra-solar planets? }

\begin{document}

\title{ Are superflares on solar analogues caused by extra-solar
planets? } 

\author{Eric P. Rubenstein\altaffilmark{1,2}} 
\affil{Yale University Astronomy Department \\
PO Box 208101, New Haven, CT, 06520-8101}
\email{ericr@astro.yale.edu}
\and 
\author{Bradley E. Schaefer}
\affil{Yale University Astronomy Department \\
PO Box 208101, New Haven, CT, 06520-8101}
\email{schaefer@grb2.physics.yale.edu}

\altaffiltext{1}{NSF Research Fellow}
\altaffiltext{2}{Present address: Cerro Tololo Inter-American
Observatory, Casilla 603, La Serena, Chile}

%\authoraddr{PO Box 208101, New Haven, CT, 06520-8101\\
%ericr@astro.yale.edu\\
%Tel \# (203) 432-3028\\
%FAX \# (203) 432-5048}

\begin{abstract}
Stellar flares with ${ 10^2-10^7}$ times more energy than the largest
solar flare have been detected from 9 normal F and G main sequence
stars (Schaefer, King \& Deliyannis 1999).  These superflares have
durations of hours to days and are visible from at least x-ray to
optical frequencies.  The absence of world-spanning aurorae in
historical records and of anomalous extinctions in the geological
record indicate that our Sun likely does not suffer superflares.  In
seeking to explain this new phenomenon, we are struck by its
similarity to large stellar flares on RS Canum Venaticorum binary
systems, which are caused by magnetic reconnection events associated
with the tangling of magnetic fields between the two stars.  The
superflare stars are certainly not of this class, although we propose
a similar flare mechanism.  That is, superflares are caused by
magnetic reconnection between fields of the primary star and a
close-in Jovian planet.  Thus, by only invoking known planetary
properties and reconnection scenarios, we can explain the energies,
durations, and spectra of superflares, as well as explain why our Sun
does not have such events.
\end{abstract}


%        ***********************************
\section{Introduction}
%        ***********************************

The discovery (Schaefer, King \& Deliyannis 1999) that some solar
analogues have large stellar flares is intriguing.  The consequences
of our Sun experiencing such a ``superflare'' (SF) would be
catastrophic for life on Earth, so it is a relief that the Sun
apparently has not had any superflares.  Still, we are strongly
motivated to determine the cause of these outbursts.  We also want to
understand if the Sun could conceivably experience such a superflare,
whether potentially life-bearing stellar systems might be affected by
these outbursts, and how to predict which systems might be subjected
to these outbursts.

The SFs occur on main sequence stars in spectral class F8 to G8 with
no unusual properties (specifically rapid rotation, high chromospheric
activity, close binary companions, or very young age).  The observed
SF energy ranges from $10^{33}$ to $10^{38}$ erg, although the
bolometric correction will in all cases substantially increase the
required total energy.  The typical SF duration is about one hour,
although the range is from a fraction of an hour up to days.  SFs emit
radiation at least from the x-rays to the optical, with indicated
temperatures (Schaefer, King \& Deliyannis 1999) from $>15000$ K
(from the HeI emission line) to $\sim 10$ keV (from x-ray continuum
fits).  Our Sun has definitely not had any SFs in the last five
centuries (due to the lack of global aurorae) or SFs with $>10^{36}$
erg in the last billion years or so (due to the lack of appropriate
extinctions in the geologic record).  Any SF model must satisfy four
constraints: the energy budget, the outburst duration, the broad range
of emitted light frequencies, and the lack of SFs on our Sun.

%        ***********************************
\section{Model by analogy}
%        ***********************************

As guidance for our model, we note that there exists a well-known
class of F and G main sequence stars with large ($\sim 10^{33}$ to
$10^{38}$ erg) flares that last from hours to days visible from x-ray
to optical frequencies (Mathioudakis et al. 1992).  This group is
named after the prototype star RS Canum Venaticorum, or RS~CVn.  These
RS~CVn systems are defined as binary stars that have an orbital period
between 1 and 14 days in which the hotter component is an F or G type
main sequence star and Ca II H\&K emission is strong at all orbital
phases (Hall 1976, 1989).  The close binary companion tidally spins up
the F or G main sequence star, which is revealed by high rotational
velocity ($\gtrsim 15$ km s$^{-1}$, Hall 1989), high chromospheric
activity (Hall 1976) (typically S$> 0.5$ Duncan et al. 1991)), high
x-ray luminosity (Strassmeier et al. 1993) ($\gtrsim 3x10^{29}$ erg
s$^{-1}$), and detectable orbital radial velocity variations (Simon,
Linsky \& Schiffer 1980).  RS~CVn flares are thought to be caused by
magnetic reconnections mediated by the close companion star (Simon,
Linsky \& Schiffer 1980 and Gunn et al., 1997).

The properties of SFs and RS~CVn flares are sufficiently close that a
similarity of mechanism is suggested.  Nevertheless, the nine SF stars
reported by Schaefer, King, and Deliyannis are definitely not RS CVn
binaries.  This is known because the SF stars have low measured
rotational velocities ($<9.7$ km/s), low Ca II H\&K emission
(S$<0.367$), low x-ray luminosities ($\lesssim 3x10^{29}$ erg s$^-1$),
and no detectable radial velocity variations (Schaefer, King \&
Deliyannis, 1999).  It appears as if the SF stars are like RS CVn
stars except that the close companion has not tidally spun-up the F or
G main sequence star.

What type of close binary companion could mediate a magnetic
reconnection (MR) event yet not spin-up the primary star?  The answer
to this question could come from the recent discovery that many F and
G main sequence stars have close planetary companions comparable or
larger in mass than Jupiter (Marcy \& Butler 1998, Butler et al. 1998).
Such Jovian planets would not spin-up their primary star, and hence
there will be no abnormal rotation velocity, chromospheric activity,
or x-ray luminosity.  The expected radial velocity variations of the
primary star will be small and unobservable without extremely careful
study.  Presumably, the Jovian planet will have a dynamo-enhanced
magnetic dipole moment comparable to or larger than that of Jupiter
which can tangle up the field of the primary star.

So we propose that the SFs occur on otherwise normal F and G main
sequence stars with a close Jovian companion, with the superflare
itself caused by magnetic reconnection in the field of the primary
star mediated by the planet.  Our proposed mechanism has the advantage
that it is known to work in a similar physical setting.  This
hypothesis also has the advantage that no exotic or rare ingredients
are required.  Indeed, the common occurrence of close Jovian planets
creating settings similar to that of RS CVn systems may perhaps
produce superflares on many normal solar-type stars.

Our proposed model for superflares on normal stars is similar to the
model proposed for large stellar flares on RS CVn binaries.  The
magnetic fields of the primary star and the planet will interact in
two ways.  First, the field lines connecting the pair will be wrapped
by orbital motion increasing the stress tensor which will manifest as
increasing magnetic field strength (Katz 1999).  Second, the
interaction of specific field loops with the passing planet will
initiate a reconnection event similar to that proposed by Simon,
Linsky \& Schiffer (1980) and Gunn et al. (1997) for large RS CVn
flares.  Presumably, the planetary motion increases the primary's
magnetic field, the interaction of magnetic loops then leads to
magnetic reconnection in the space between the star and planet, Alfven
waves are generated which propagate toward the star, and the magnetic
energy in the Alfven waves accelerates particles near the stellar
surface to emit x-rays and optical light (see Haisch, Strong \& Rodono
1991 for a review of the related physics of solar flares).
Alternatively, the reconnection might occur near the surface of the
star where the higher particle density leads to a lower Alfven
velocity (Lazarian 1999).

Indeed, this latter case is likely what happened for the $\kappa$ Ceti
superflare.  Robinson \& Bopp's (1987) observation of the $5876 \AA $
He D$_3$ line in emission (the transition between $1s2p^{3}P$
and $1s3d^{3}D$ states) indicates an outburst at or close to the
stellar photosphere.  This line is observed in emission from
significant solar flares (type 2 or greater) when the plasma density
is n$_{e^{-}} > 10^{14} cm^{-3}$ (Feldman, Liggett \& Zirin 1983).
Solar models (Allen 1973) indicate that this is the electron
density essentially at the photosphere. 

Unfortunately, it is difficult to be specific with regards to the
physical mechanism of the explosion since we still do not have a
detailed understanding of MR energy release mechanisms despite active
research into magnetic reconnection events in laboratory plasmas (Ji
et al. 1998, Wesson et al. 1997, Yamada et al. 1997, Ono et al. 1996)
and terrestial/planetary magnetospheres (Russell et al. 1998, Russell
et al. 1997).  The physics of the magnetic reconnection in
well-observed solar flares (Antiochos, Devore \& Klimchuk 1999, Somov
\& Kosugi 1997, Haisch, Strong \& Rodono 1991) and RS~CVn systems
(Ferreira 1998) are also only crudely known.  There are still
significant differences between competing theoretical models of how MR
unfolds (Lazarian \& Vishniac 1999, Hanato et al. 1998, Biskamp 1997,
Pellat, Hurricane \& Luciani 1996, Biskamp, Schwarz \& Drake 1995).
So any SF MR model cannot provide a unique physical explanation of the
magnetic reconnection mechanism.

Despite the uncertainties in the detailed physics, we do know that
magnetic reconnection occurs in physical settings similar to those of
a G-star with a magnetized companion.  Fortunately, even without
knowledge of the detailed physics, we can still make estimates of the
energy budget, the burst duration, and the flare spectrum.

The total magnetic energy outside the star will be equal to B$^2/8\pi$
times the volume of the star for a dipole field, with B equal to the
field strength at the surface of the star.  For total annihilation of
the field outside a solar-size star, the required B varies from 6 to
$1200$ Gauss for the observed range of SF energy (from $2x10^{33}$ to
$9x10^{37}$ erg).  However, the energy in a dipole configuration will
not reconnect to provide available energy.  In general, only some
fraction, f, of the magnetic field will be in higher order moments
susceptible to reconnection.  With this fraction, the available energy
for the flare then becomes $E=(B^{2}/8\pi )(4\pi R^{3}/3)f$.  If f is
large, say 0.1, as possible for significant flux caused by the winding
of the primary's field, then the associated field strength for a SF
would range from roughly $10$ to $3000$ Gauss.  If f is relatively
small, say $0.001$, as appropriate for a large spot, then the surface
field strength will depend on the exact structure of the field,
although the above formula would suggest a field strength of order
$100$ to $30000$ Gauss.  With individual spots on our Sun having
characteristic fields of up to $3000$ Gauss, these required magnetic
field energies do not seem unreasonable for SF stars.  The observed
magnetic energy on our Sun would be sufficient to power a SF, if only
a mechanism (such as interactions with a nearby planetary dipole)
would allow for reconnection.  For the most energetic superflares or
to allow for inefficiencies in the conversion of magnetic to radiative
energy, we expect that the surface magnetic fields on SF stars might
be larger than the values quoted above.

So we are predicting that SF stars likely have magnetic fields
substantially higher than on our own Sun.  The measure of magnetic
field strength on solar-type stars is not easy and is best performed
only on bright stars.  To date, only two of the nine known SF stars
have measured magnetic field strengths.  $\kappa$ Ceti has a magnetic
field strength of $1500$ Gauss over about $35$\% of its surface
(Montesinos \& Jordan 1993).  $\pi^{1}$ UMa has an average surface
field of $1900$ Gauss (Gray 1984).  These fields imply total magnetic
energies of $\approx 2x10^{37}$ and $2x10^{38}$ ergs respectively.
With observed SF energies of $2x10^{34}$ and $2x10^{33}$ ergs, the two
SFs used only $10^{-3}$ and $10^{-5}$ of the magnetic energy available
on the stars respectively.  Since it is possible that the heating of
the plasma by Alfven waves may be highly inefficient (Lazarian \&
Vishniac 1999) we expect that typically only a small fraction of the
available magnetic energy is observed in SFs.  However, these
fractions are sufficiently low that it is plausible that the stars'
magnetic fields will have configurations with this much energy
available for reconnection.  Thus, we take the measured high magnetic
field on two-out-of-two SF stars to represent a successful prediction
of our model as well as to demonstrate that the energy available
within our model accounts for the observed energetics of superflares.

Gray (1984) notes that $\pi^{1}$ UMa is unique in his sample of being
an early type star with a high magnetic field.  This unusual presence
of a high field provides a distinct peculiarity between SF stars and
other normal solar-type stars.  The existence of this peculiarity on a
SF star also suggests a connection between the high magnetic field and
the SF.  Our model provides a causal connection between the high
magnetic field and the superflare.

What does our model predict for the typical time scales in the SF
light curves?  One way to make the prediction is by analogy with the
RS CVn stars which have time scales from hours to days (Osten \& Brown
1999).  A second way is to realize that the typical rise times will be
of order the time it takes for an Alfven wave to cross the primary
emission region.  For reconnection events midway between a planet and
the star, the characteristic size scale will be $\approx 0.1$ AU.
However, the bulk of the emission is likely to arise when the Alfven
waves encounter dense gas near the star's surface.  The characteristic
size scale would then be more like a solar radius ($7x10^{10}$ cm).
The Alfven velocity can be estimated in the case of the SF on $\kappa$
Ceti, where the density of the emitting region was $1.6x10^{-10}$
gm/cm$^{3}$ (Robinson \& Bopp 1988) and the magnetic field is $1500$
Gauss (Montesinos \& Jordan 1993), so that the Alfven velocity in the
emission region is near $3.3x10^{7}$ cm/s$^{1}$.  The crossing time is
thus of order 30 minutes, which is comparable to the time scale for
the observed superflare on $\kappa$ Ceti.  For weaker magnetic fields,
smaller emission regions, or propagation at slower than the Alfven
velocity, the rise times and durations can be substantially larger
than 30 minutes. Thus, our model predicts rise times and durations
from hours to days for SFs, and this is just as observed.

The Alfven waves will accelerate particles to a wide range of
energies, with a spectrum that is difficult to predict from first
principles.  Nevertheless, we expect that SFs will have spectra
comparable to those of RS CVn flares and also of solar flares.  Thus,
we expect that SFs will emit large amounts of x-rays, optical light,
and even radio emission.  As with solar flares (Foukal 1990; Robinson
\& Bopp 1988), regions above the star's surface will be heated to
temperatures of $>15000$ K to form prominent He I emission lines.  So
our expected broad range of energies is matched by that observed for
superflares, with a prediction that SFs will also be bright at radio
frequencies.

This model shares with RS CVn flares all of the essential physics.
The substitution of a Jovian planet for the RS CVn's non- or
less-active component should not alter these properties since the role
of this companion is merely to anchor a magnetic field.  As long as
the companion has a magnetic dipole moment of adequate strength, the
composition and physical size of the companion is not important.
Therefore, the energy budget, outburst time-scale, and range of
emitted frequencies should be comparable.  Thus, we take the RS CVn
flares as proof that our SF model satisfies the first three
constraints.  The fourth constraint, the lack of SFs on our Sun, is
also satisfied since our Solar System does not have a planet with a
large magnetic dipole moment in a close orbit.

%        ***********************************
\section{Discussion}
%        ***********************************

Our SF model makes two testable predictions.  First, we require that
SF stars have close-in planets that might be revealed by small radial
velocity variations of the primary star.  The masses and orbital
periods of the planets are not well constrained by our model and the
orbital inclination might be near face-on, so some SF stars might
easily not have detectable planets by this method.  Only one SF star
($\kappa$ Ceti) has been searched for planets to date, and its radial
velocity variations (with a 20 m/s dispersion) allow for a Saturn-mass
planet even at low inclination (Marcy private communication).  Second,
the SF stars must have a sufficient magnetic field to provide the
observed flare energy.  This would require surface dipole fields of
$\sim 6-1200$ Gauss should the entire field be annihilated, and
proportionally more should only a fraction be annihilated.  This
prediction has already been partially confirmed by the high magnetic
fields on the only two SF stars tested to date.

Within our model, SFs will occur only on solar-type stars with
planetary systems, and these are the same systems in which life can
form.  These planets will be close to the primary, where the flux from
the SF will be large and might have profound consequences for
formation and survival of extra-terrestrial life; such life might be
fundamentally different from terrestial life.  These consequences
might include the provision of energy needed to initiate biological
processes or the destruction of life forms by high radiation levels.

\acknowledgments

We would like to thank an anonymous referee for helpful suggestions
for the preparation of this letter.  EPR acknowledges the support of
NSF grant INT-9902667.

\end{document}